\documentstyle[aps,pre]{revtex}

\begin{document}
\draft
\title{Spatiotemporal chaotic dynamics of solitons with internal structure 
in the presence of finite-width inhomogeneities}
\author{L. E. Guerrero$^1$, A. Bellor\'\i n$^2$, J. R. Carb\'o$^3$ and J. A.
Gonz\'alez$^4$}
\address{$^1$Departamento de F\'{\i}sica, Universidad Sim\'on Bol\'{\i}var\\
Apartado 89000, Caracas 1080-A, Venezuela\\
$^2$Departamento de F\'\i sica, Facultad de Ciencias\\
Universidad Central de Venezuela\\
Apartado Postal 47588 Los Chaguaramos, Caracas 1041-A, Venezuela\\
$^3$Departamento de F\'\i sica, Universidad de Camag\"uey, Circunvalaci\'on\\
Norte, Camag\"uey 74650, Cuba\\
$^4$Centro de F\'\i sica\\
Instituto Venezolano de Investigaciones Cient\'\i ficas\\
Apartado Postal 21827, Caracas 1020-A, Venezuela} 
\date{\today }
\maketitle

\begin{abstract}
We present an analytical and numerical study of the Klein-Gordon
kink-soliton dynamics in inhomogeneous media. In particular, we study an
external field that is almost constant for the whole system but that changes
its sign at the center of coordinates and a localized impurity with
finite-width. The soliton solution of the Klein-Gordon-like equations is
usually treated as a structureless point-like particle. A richer dynamics is
unveiled when the extended character of the soliton is taken into account.
We show that interesting spatiotemporal phenomena appear when the structure
of the soliton interacts with finite-width inhomogeneities. We solve an
inverse problem in order to have external perturbations which are generic
and topologically equivalent to well-known bifurcation models and such that
the stability problem can be solved exactly. We also show the different
quasiperiodic and chaotic motions the soliton undergoes as a time-dependent
force pumps energy into the traslational mode of the kink and relate these
dynamics with the excitation of the shape modes of the soliton.
\end{abstract}


\preprint{HEP/123-qed}

\section{Introduction}

For a variety of systems the interplay between nonlinearity and disorder
results in novel and fascinating phenomena \cite{j:1,j:2}. Particularly, the
study of soliton dynamics in inhomogeneous and disordered media has received
a great deal of attention in recent years \cite
{j:3,j:4,j:5,j:6,j:7,j:8,j:9,j:10,j:11,j:12,j:13,j:14} since it concerns
real condensed matter systems and phenomena.

It is well known that transport properties in inhomogeneous and disordered
media can change dramatically when the nonlinearity allows the creation of
solitons. As a first step in the study of soliton dynamics in disordered
media many authors explored the interaction between a soliton and an
isolated impurity. This soliton-impurity interaction has been modeled by a
point-like impurity and a structureless soliton. In real situations we can
have several inhomogeneities of different kinds. When the distance between
impurities is considerable higher than the width of solitons and impurities
the traditional approach gives a correct result. When this is not so, we
witness a series of surprising phenomena.

The dimensionless Klein-Gordon-like equations model a wide variety of
soliton bearing systems\cite
{j:1,j:2,j:11,hanggi,jj:1,jj:2,jj:3,jj:4,jj:5,jj:6,jj:7,jj:8,jj:9,jj:10,jj:15,jj:16,jj:20,j:40,j:41,j:50}%
, including charge density waves, Josephson junctions, structural phase
transitions, crystal growth, polymers, proton conductivity, macromolecules
and hydrogen-bond chains: 
\begin{equation}
\phi _{xx}-\phi _{tt}-\frac{dU}{d\phi }={\cal F}(\phi ,\phi _t,x,t);
\label{equ_1}
\end{equation}
here $U=U(\phi )$ is a potential that possesses at least two minima\cite
{j:61}, meanwhile ${\cal F}(\phi ,\phi _t,x,t)$ represent additional forces
(external forcing, dissipation, presence of impurities, inhomogeneous
external fields, coupling to other degrees of freedom).

The soliton solution of Eq. (\ref{equ_1}) is usually treated as a
structureless point-like particle. A richer dynamics is unveiled when the
extended character of the soliton is taken into account. For instance, in a
previous work \cite{j:6}, Gonz\'alez and Ho\l yst studied the $\phi ^4$
equation ($U=\frac 18(\phi ^2-1)^2$): 
\begin{equation}
\phi _{xx}-\phi _{tt}-\gamma \phi _t+\frac 12\left( \phi -\phi ^3\right)
=-F(x)-G(x,t).  \label{e:1}
\end{equation}
Particularly, they showed that the zeroes of $F(x)$ (for $G\equiv 0$) play
the roll of equilibrium positions for the soliton. In the case that only one
zero, $x_0$ exists, the condition for stability for the kink/antikink is: 
\begin{equation}
\left[ \frac{dF}{dx}\right] _{x=x_0}\left\{ 
\begin{array}{ll}
>0 & \mbox{for the kink.} \\ 
<0 & \mbox{for the antikink.}
\end{array}
\right.  \label{con:1}
\end{equation}
For the stable case, the inhomogeneity can trap the soliton, but in the case
that $F(x)$ possesses more than one zero, the stability condition may become
too complex. Here the extended character of the soliton arises, and
interesting phenomena appear such as the interaction between the structure
of the soliton (which is not a point-like particle) and the inhomogeneities,
and between the shape modes of the soliton themselves. The external force $%
F(x)$ can change the spectrum of small oscillations about the soliton and
additional bounded states can exist.

For the soliton of the $\phi ^4$ equation without perturbation, there are
only two bounded states (besides the continuous spectrum): the translational
mode and the shape mode. For specific values of the parameters that define
the force $F(x)$ used in Ref. \cite{j:6}, not only an increase of the number
of shape modes can exist, but in certain cases these can be unstable.
Moreover the continuous spectrum can lose stability and the soliton becomes
unstable against interaction with phonons.

The above mentioned authors also considered the problem with the time
dependent force $G(x,t)$ and found that if $G(x,t)$ has a spatial shape such
that it coincides with one of the eigenfunctions of the stability operator
of the soliton, then it is possible to get resonance if the frequency of the
force also coincides with the resonant frequency of the considered mode.
This means, for example, that energy can be given to the translational mode
(or any of the shape modes) using a $G(x,t)$ coupled to the translational
mode (or any of the shape modes).

When $F(x)$ has three zeroes, this is equivalent to a double well potential
(like the one found in the Duffing equation\cite{j:6,j:15}) chaotic motion
of the soliton is possible applying an additional periodic force $G(x,t)$
for a determined set of values of the parameters.

In this paper we take into account the extended character of both the
soliton and the impurity. We show that these considerations lead to the
existence of a finite number of soliton internal modes that underlies a rich
spatiotemporal dynamics.

We introduce impurities of the $N(x)\phi $ type, where $N(x)$ is a function
with a bell shape. An impurity of this kind but using delta functions has
been presented in Ref. \cite{j:14}. In our paper we consider a finite width
impurity and a finite width kink and show the striking differences between
this approach and the traditional one (structureless solitons and
delta-function-like impurities).

We present a model for which the exact stationary soliton solution in the
presence of inhomogeneities can be obtained and the stability problem can be
solved exactly. To achieve this purpose we solve an inverse problem in order
to have external perturbations which are generic and topologically
equivalent to well-known bifurcation model systems\cite{j:51}.

We choose the ``exact'' solution such that the differential operator that
appears in the stability problem is a Posch-Teller potential that can be
solved exactly. Besides, the ``generated'' external force and the impurities
have important physical properties. In particular, the inhomogeneous force $%
F(x)$ is equivalent to the pitchfork bifurcation canonical form\cite{j:15}
and the $N(x)\phi $ impurity is topologically equivalent to the $\delta
(x)\phi $ type impurity that is very frequently used.

Furthermore, we demonstrate the sensibility of the soliton internal dynamics
to the inhomogeneity width even for the isolated impurity case.

Our paper is organized as follows. In Sec. II we present a description of
our model and we give specific physical interpretations of the equations
under consideration. In Sec. III we study the equilibrium positions of the
soliton and its stability. We analyze the interaction of the soliton with
the inhomogeneity created by the interplay between a finite-width and $\phi $%
-dependent impurity and the already studied inhomogeneities independent of $%
\phi $ \cite{j:6}. We also consider the action of time dependent forces
fitted to the shape of the translational mode of the soliton. In Sec. IV we
describe the numerical simulations that confirm the theoretical results. We
use the Karhunen-Lo\`eve decomposition to relate the excitation of the shape
modes spectrum of the soliton with the bifurcations. In Sec. V we present
the interaction of the soliton with radiation modes. Finally, in Sec. VI we
summarize and discuss our results and also present some concluding remarks.
In the Appendices we outline the numerical method and present the
Karhunen-Lo\`eve decomposition.

\section{The model}

The topological solitons studied in the present paper possess important
applications in condensed matter physics. For instance, in solid state
physics, they describe domain walls in ferromagnets or ferroelectric
materials, dislocations in crystals, charge-density waves, interphase
boundaries in metal alloys, fluxons in long Josephson junctions and
Josephson transmission lines, etc.\cite{j:11,jj:7}

Although some of the above mentioned systems are described by the $\phi ^4$%
-model and others by the sine-Gordon equation (and these equations, in their
unperturbed versions, present differences like the fact that the sine-Gordon
equation is completely integrable whereas the $\phi ^4$-model is not) the
properties of the solitons supported by sine-Gordon and $\phi ^4$ equations
are very similar. In fact, these equations are {\it topologically equivalent}
and very often the result obtained for one of them can be applied to the
other\cite{jj:7}.

Here we consider the $\phi ^4$ equation in the presence of inhomogeneities
and damping:

\begin{equation}
\phi _{xx}-\phi _{tt}-\gamma \phi _t+\frac 12\left( \phi -\phi ^3\right)
=-N(x)\phi -F(x),  \label{a:1}
\end{equation}
where $F(x)$ is a function with (at least) one zero and $N(x)$ is a
bell-shaped function that rapidly decays to zero for $x\rightarrow \pm
\infty $.

In ferroelectric materials $\phi $ is the displacement of the ions from
their equilibrium position in the lattice, $\frac 12\left( \phi -\phi
^3\right) $ is the force due to the anharmonic crystalline potential, $F(x)$
is an applied electric field, and $N(x)$ describes an impurity in one of the
anharmonic oscillators of the lattice\cite{s:3}. In Josephson junctions, $%
\phi $ is the phase difference of the superconducting electrons across the
junction, $F(x)$ is the external current, and $N(x)$ can describe a
microshort or a microresistor\cite{j:5}. In a Josephson transmission line it
is possible to apply nonuniformly distributed current sources ($F(x)$) and
to create inhomogeneities of type $N(x)$ using different electronic circuits
in some specific elements of the chain\cite{j:11,s:5}.

In the present paper the functions $F(x)$ and $N(x)$ will be defined as,

\begin{equation}
F(x)=\frac 12A(A^2-1)\tanh (Bx),  \label{a:10}
\end{equation}

\begin{equation}
\ N(x)=\frac 12\frac{(4B^2-A^2)}{\cosh {}^2(Bx)}.  \label{a:11}
\end{equation}

The case $F=const.$ has been studied in many papers (see e.g. \cite{jj:7}).
Here Eq.~(\ref{a:10}) represents an external field (or a source current in a
Josephson junction) that is almost constant in most part of the chain but
changes its sign in $x=0$ (this is very important in order to have soliton
pinning\cite{j:6}). Microshorts, microresistors and/or impurities in atomic
chains\cite{j:5} are usually described by Dirac's delta functions ($\delta
(x)$) where the width of the impurity is neglected. The function $N(x)$ is
topologically equivalent to a $\delta (x)$ but it allows us to consider the
influence of the width of the impurity.

\section{Stability Analysis}

\label{sec2}

Let us consider the Eq. (\ref{a:1}) and assume the existence of a static
kink solution $\phi _k(x)$ that corresponds to a soliton placed in a stable
equilibrium state created by the inhomogeneities $F(x)$ and $N(x)$.

Analyzing the small amplitude oscillations around the kink solution $\phi
_k(x)$,

\begin{equation}  \label{a:2}
\phi (x,t)=\phi _k(x)+\psi (x,t),
\end{equation}

we get, for the function $\psi (x,t)$, the following equation:

\begin{equation}
\psi _{xx}-\psi _{tt}-\gamma \psi _t+\frac 12(1-3\phi _k^2+2N(x))\psi =0.
\label{a:3}
\end{equation}

Studying of the stability of the equilibrium solution $\phi _k(x)$ leads to
the following eigenvalue problem (we have introduced $\psi (x,t)=f(x)\exp
(\lambda t)$ into Eq.~(\ref{a:3})):

\begin{equation}
-f_{xx}+\frac 12(3\phi _k^2-1-2N(x))f=\Gamma f,  \label{a:4}
\end{equation}
where $\Gamma \equiv -\lambda ^2-\gamma \lambda $.

Let us now study some particular cases: If $F(x)\equiv 0$, and the function $%
N(x)$ is described by the expression (\ref{a:11}), then it can be shown that
the exact solution for the kink in equilibrium at the position $x=0$ is,

\begin{equation}  \label{a:6}
\phi _k(x)=\tanh (Bx),
\end{equation}

and that the discrete eigenvalue spectrum is described by the following
formula,

\begin{equation}  \label{a:7}
\Gamma _n=B^2(\Lambda +2\Lambda n-n^2-2);
\end{equation}
where the parameter $\Lambda $ is defined as,

\begin{equation}  \label{a:8}
\Lambda (\Lambda +1)=\frac 1{B^2}+2.
\end{equation}

The integer part of $\Lambda $ ($\left[ \Lambda \right] $) defines the
number of modes of the discrete spectrum.

From Eq.~(\ref{a:7}) the stability condition for the translational mode ($%
n=0 $) can be obtained:

\begin{equation}  \label{a:9}
4B^2<1.
\end{equation}

It is worth noticing that if the coefficient of $N(x)$ is negative, the
equilibrium position created by the impurity of the $N(x)\phi $ type is
stable for the soliton, and that the stability condition is independent of
the polarity of the soliton. The opposite occurs when a inhomogeneity like $%
F(x)$ is considered.

Furthermore, it is necessary to point out the differences between the case
in which the soliton equilibrium position is due to a zero of $F(x)$, and
the case in which the kink is trapped in the effective well created by the
impurity $N(x)\phi$. In the former the characteristic width and the number
of internal modes that can be excited are smaller than those for the free
kink, and even smaller than those for the kink in equilibrium in an unstable
position. While, in the latter the characteristic width and the number of
internal modes that can be excited are greater than those for the free kink
and for the kink in unstable equilibrium in a repulsive impurity. This is
due to the inverse proportionality between the width of the impurity and $B$
(just like the width of the kink). The impurity is more stable as $B$
diminishes.

Finally, let us consider the case in which both types of inhomogeneities ($%
F(x)$ and $N(x)\phi $) are present in Eq.~(\ref{a:1}). These functions are
defined as in equations (\ref{a:10}) and (\ref{a:11}).

For these functions the exact solution describing the static soliton can be
written:

\begin{equation}  \label{a:12}
\phi _k(x)=A\tanh (Bx).
\end{equation}

The spectral problem (Eq.~(\ref{a:4})) brings the following eigenvalues for
the discrete spectrum:

\begin{equation}  \label{a:13}
\Gamma _n=\frac 12A^2-\frac 12+B^2(\Lambda +2\Lambda n-n^2-2);
\end{equation}
here $\Lambda $ is defined as,

\begin{equation}  \label{a:14}
\Lambda (\Lambda +1)=\frac{A^2}{B^2}+2.
\end{equation}

The stability condition for the translational mode is,

\begin{equation}
16B^4+2B^2(5-7A^2)+(1-A^2)^2<0.  \label{a:15}
\end{equation}

When this condition is not fulfilled (the equilibrium position $x=0$ is
unstable) and $A^2>1$, then there will exist three equilibrium points for
the soliton: two stable (at points $x=x_1>0$ and $x=x_2<0$) and one unstable
at point $x=0$. This happens because for large values of $\left| x\right| $
the leading inhomogeneity is $F(x)$, which is non-local and not zero at
infinity. This inhomogeneity acts as a restoring force that pushes the
soliton towards the point $x=0$. As a result of the competition between the
local instability induced by $N(x)\phi$ at $x=0$ and the non-local
inhomogeneity $F(x)$, an effective double-well potential is created. This is
equivalent to a pitchfork bifurcation.

Also note that when $A=0$ then $\Lambda =1$, i.e., there is an unstable mode
created by the impurity $N(x)\phi $ even for a flat initial condition. This
contrasts with the $\Lambda =0$ result expected for a delta-like impurity
and with the result of Gonz\'alez and Ho\l yst\cite{j:6} for the other kind
of inhomogeneity ($F(x)$).

\section{Quasiperiodic and chaotic solitons}

In this section we present numerical results for Eq. (\ref{equ_1}) perturbed
with inhomogeneous external forces, impurities and time-periodic forces.
First, we show, for inhomogeneities of the $F(x)$ type, the bifurcations
leading to the chaotic regime. Later, we study quasiperiodic regimes as the
soliton internal modes are excited by an impurity of the type $N(x)\phi $.
Finally, we present bifurcations as the internal modes are excited for the
case in which the two types of inhomogeneities ($F(x)$ and $N(x)\phi $) are
present.

\subsection{Scenario of inhomogeneous external forces}

We simulate the following equation,

\begin{equation}
\phi _{xx}-\phi _{tt}-\gamma \phi _t+\frac 12\left( \phi -\phi ^3\right)
=-F(x)-G(x,t),  \label{b:1}
\end{equation}
where static and time-dependent forces are inhomogeneous and given by the
following expressions:

\begin{equation}
F(x)=\frac 12A(A^2-1)\tanh (Bx)+\frac 12A(4B^2-A^2)\sinh (Bx)\cosh
{}^{-3}(Bx)  \label{b:2}
\end{equation}
and 
\begin{equation}
G(x,t)=\nu \cos (\omega t)\left( \frac 1{\cosh {}^2(E(x-x_1))}+\frac 1{\cosh
{}^2(E(x+x_1))}\right).  \label{b:3}
\end{equation}

The force $F(x)$, firstly introduced in Ref. \cite{j:6}, can be obtained as
a result of an inverse problem for which an effective potential equivalent
to the canonical form of the pitchfork bifurcation is desired. For $A^2>1$
and $B^2\left( \sqrt{1+\frac{6A^2}{B^2}}-1\right) <1$, we are in presence of
a double-well potential {\it for the soliton}, (i.e., the model is a
Duffing-like\cite{j:15} soliton oscillator). Note that a force with three
zeroes does not imply necessarily the existence of a double-well potential
for the soliton{\it .}

We will attend to the dynamics of the center of mass as defined by Eq. (\ref
{aa:3}). Figure~\ref{Fig.1} presents the bifurcation diagram ($A=\sqrt{\frac 
32}$, $B=\sqrt{\frac 1{10}}$, $\Lambda =2$, $\omega =1.22$, $E=0.1$, $%
x_1=2.5 $ and $\gamma =0.450$) as the amplitude $\nu $ of the time-dependent
driving force is increased. The period doubling cascade for low values of
the control parameter corresponds to oscillations of the soliton in one
well. At $\nu =0.312$ an unusual discontinuous transition interrupts the
cascade: the system switches from a period-eight solution to a period-two
solution but in the other well as can be appreciated in the time series
presented in figures \ref{Fig.2}(a)-(b). At $\nu =0.318$ the period-eight
solution is recovered and a two-band regime with n-periodic (Fig. \ref{Fig.3}%
(a)), quasiperiodic (as the two-torus attractor presented in Fig. \ref{Fig.3}%
(b)) and chaotic attractors. The quasiperiodic attractor is a spatiotemporal
effect: it is generated by the activation of a shape mode of the soliton
which provides a frequency that is incommensurate to the driving frequency.
Figures \ref{Fig.1} and \ref{Fig.2}(c) reveals for $\nu =0.332$ the onset of
jumps of the soliton between the two wells. This regime corresponds to
Duffing-like chaos for the soliton. Figure \ref{Fig.3}(b) also presents the
Poincar\'e map which reveals the high-dimensional chaotic motion of the
soliton which can be ascribed to an increased activation of shape modes
(which is not possible for a Duffing-like chaotic particle). The intense
activity around $x_{c.m.}=0 $ (Fig. \ref{Fig.1}) is also due to the extended
character of the soliton. We have verified that at $\nu =0.356$ the soliton
prefers its deformation instead of its destruction and its dynamics returns
to a periodic solution (figures \ref{Fig.1} and \ref{Fig.2}(d)).

Figures \ref{Fig.4}(a)-(b) present time series for the center of mass of
chaotic solitons. In the first case ($\nu =0.3$ and $\gamma =0.505$, the
rest parameters are preserved) the soliton jumps between two wells. When the
damping is increased (Fig. \ref{Fig.4}(b), $\gamma =0.550$), the soliton is
constrained to move mainly in one well. Corresponding Poincar\'e maps (Fig. 
\ref{Fig.5}(a)-(b)) show the contrast of dimensionality (and consequently,
of the number of modes and/or of the effective number of degrees of freedom)
of these chaotic motions. Figures \ref{Fig.5}(c)-(d) present the
spatiotemporal evolution for these chaotic solitons. Figure \ref{Fig.5}(c)
evidences the jumps of the domain wall between the two wells, whereas Fig. 
\ref{Fig.5}(d) shows such a domain wall oscillating in one of the wells.
Notwithstanding, note also in Fig. \ref{Fig.5}(d) the appearance and
disappearance of deformations (depicted by yellow zones) in the other well
due to tunnelling of mass/energy.

\subsection{Scenario of finite-width impurity}

We simulate the following model,

\begin{equation}
\phi _{xx}-\phi _{tt}-\gamma \phi _t+\frac 12\left( \phi -\phi ^3\right)
=-N(x)\phi -G(x,t),  \label{b:4}
\end{equation}
where external forces are time-dependent and inhomogeneous. Right-hand
functions are given by,

\begin{equation}
N(x)=\frac{\frac 12\left( 4B^2-1\right) }{\cosh {}^2(Bx)}  \label{b:5}
\end{equation}
and 
\begin{equation}
G(x,t)=\frac{\nu \cos (\omega t)}{\cosh {}^\Lambda (Bx)}.  \label{b:6}
\end{equation}

One may think that soliton dynamics in presence of an attractive impurity
must be simple. Notwithstanding, that is not the case: as has been discussed
in Section~\ref{sec2} when the width of the impurity is finite, a large
number of internal modes can be excited. It can exist energy exchange
between these modes and the translational mode bringing a complex dynamics.

Figure \ref{Fig.6}(a) shows the phase space for a period-three solution ($%
A=1.0$, $B=0.25$, $\Lambda =3.772$, $\omega =1.0$, $\nu =0.16$ and $\gamma
=0.1$). As the amplitude $\nu $ of the time-dependent driving force is
increased, the excitation of an internal mode can provide a frequency
incommensurate to the driving frequency, generating a two-torus
quasiperiodic attractor (Fig. \ref{Fig.6}(b) for $\nu =0.195$, the rest
parameters are the same as the previous case). On the other hand, decreasing
of the damping parameter $\gamma $ can provide even more incommensurate
frequencies to the driving frequency as reveals the three-torus
quasiperiodic motion presented in figures \ref{Fig.6}(c)-(d) for $\gamma
=0.01$ and $\upsilon =0.12$.

\subsection{Scenario of finite-width impurity and inhomogeneous forces}

We will consider the following model:

\begin{equation}
\phi _{xx}-\phi _{tt}-\gamma \phi _t+\frac 12\left( \phi -\phi ^3\right)
=-N(x)\phi -F(x)-G(x,t),  \label{b:7}
\end{equation}
where $N(x)$ and $F(x)$ are given by Eq.~(\ref{a:10}) and Eq.~(\ref{a:11}),
and $G(x,t)$ is given by the following expressions,

\begin{equation}
G(x,t)=\nu \cos (\omega t)\left( \frac 1{\cosh {}^\Lambda (B(x-x_1))}+\frac 1%
{\cosh {}^\Lambda (B(x+x_1))}\right).  \label{b:9}
\end{equation}

We will study the sequence of bifurcations as the driving amplitude $\nu $
is increased and other parameters remain fixed ($A=1.22$, $B=0.32$, $\Lambda
=2$, $\omega =1.22$, $x_1=2.5$ and $\gamma =0.3)$. Figure \ref{Fig.7}
presents the time-averaged spatial profile for different values of the
amplitude $\nu $ of the time-dependent driving force: $\nu =0.20$ (period
one), $\nu =0.28...0.45$ (quasiperiodicity), $\nu =0.55$ (chaos) and $\nu
=0.60$ (period one).

Figures \ref{Fig.8}(a)-(d) present a sequence of ``quasiperiodic
bifurcations'': torus entangles as the amplitude of the time-dependent
driving force increases and shape modes are activated.

Figures \ref{Fig.9}(a)-(b) present an chaotic soliton for $\nu =0.55$. The
Poincar\'e map (Fig. \ref{Fig.9}(a)) reveals the high-dimensional chaotic
motion due to the activation of many internal modes whereas Fig. \ref{Fig.9}%
(a) presents the temporal evolution of the soliton and shows that the kink
profile is still sustained. For these parameters the soliton is at the edge
of its destruction due to the activation of the shape modes. We can consider
this regime as fully developed shape chaos for the soliton.

Along this paper we had emphasized the importance of the internal modes (or
shape modes) of the soliton that can be excited as the soliton interacts
with inhomogeneities. Therefore we perform a Karhunen-Lo\`eve decomposition
for the sequence of dynamic attractors already presented for the scenario of
finite impurity and homogeneous force.

Figure \ref{Fig.10}(a) reveals the increasing excitation of the discrete
internal modes as the system evolves into a chaotic regime as well as the
sudden change of the spectra for the final state that correspond to solution
in which periodic motion is regained ($\nu =0.60$). The periodic solution
for $\nu =0.60$ corresponds to the highest deformity of the kink profile
(Fig. \ref{Fig.7}). This agrees with the higher contribution to the dynamics
of the first two modes whereas all the rest of the modes decreased their
contribution. Furthermore, the first shape mode replaces the translational
mode as the leading mode of the dynamics. Figure \ref{Fig.10}(b) presents
the leading Karhunen-Lo\`eve eigenmodes for the period-one solutions that
initiates and ends the sequence of bifurcations considered in this section.
The eigenmode for $\nu =0.20$ appears to be the superposition of a pair of
translational modes centered at the equilibrium points for the soliton.
Similar situation occurs for $\nu =0.60$ but the eigenvalue appears to be
the superposition of a pair of shape modes. Figures \ref{Fig.11}(a)-(b) the
striking difference of the temporal evolution of the period-one solitons for 
$\nu =0.20$ and $\nu =0.60$.

\section{Interaction of the Soliton with Radiative Modes}

Any process that involves inelastic interactions or accelerations of the
soliton leads to the emission of quasi-linear waves (radiation). This
phenomenon occurs by means of the modes of the continuous spectrum
(radiation modes\cite{j:40}) that for the case of the unperturbed $\phi ^4$%
-equation are given by

\begin{equation}
f_k(x)=e^{ikx}\left[ 3\tanh ^2\left( \frac x2\right) -6ik\tanh \left( \frac x%
2\right) -\left( 1+4k^2\right) \right] .  \label{cc:1}
\end{equation}

The interaction of a soliton with an inhomogeneity results in an emission of
radiation that can be calculated using the method of McLaughlin and Scott%
\cite{j:40} that relies on the construction of a Green's function that
consists of a bilinear combination of eigenfunctions (i.e., in our case, of
the eigenfunctions that we have already presented). The radiation problem
also has been addressed using other perturbative methods\cite{j:5}.

In this paper we have focused our attention on the stability of the
translational mode. Note that when the soliton is in an equilibrium position
created by the impurities that is stable for the translational mode, the
soliton as a whole is stable against the emission or absortion of radiation
(these modes are usually called phonon modes). Moreover, in this case the
kink is oscillating in an effective potential well for which radiative
effects are exponentially small\cite{j:5}.

Notwithstanding, we remark that when the equilibrium position is unstable
the shape modes can also become unstable (under certain conditions) and this
can destroy the soliton. Moreover, under a certain condition the modes of
the continuous spectrum can also be unstable. This is very surprising as it
leads the soliton to be unstable against the emission and the absortion of
radiation modes. This interesting effect is shown in Figure \ref{Fig.12}
where we present the interaction of a radiation mode with the soliton under
instability conditions for the continuous spectrum. The usual methods for
calculating the radiation can not give this result. For instance, the
phenomenon shown in Figure \ref{Fig.12} is produced by the evolution of the
system

\begin{equation}
\phi _{xx}-\phi _{tt}-\gamma \phi _t+\frac 12\left( \phi -\phi ^3\right)
=-F(x),  \label{cc:2}
\end{equation}

\begin{equation}
F(x)=B(4B^2-1)\tanh (Bx)  \label{cc:3}
\end{equation}
under initial conditions that represent the supperposition of a radiation
mode and a soliton,

\begin{equation}
\phi (x,0)=2B\tanh (Bx)+C\left\{ 3\sin (Bx)\tanh (Bx)-6\cos (Bx)\tanh
(Bx)-5\sin (Bx)\right\} ,  \label{cc:4}
\end{equation}

\begin{equation}
\phi _t(x,0)=0  \label{cc:5}
\end{equation}

For $4B^2-1<0$ the translational mode is unstable. For $10B^2-1<0$ the first
shape mode also becomes unstable. Moreover, for $12B^2-1<0$ the soliton
becomes unstable against the interaction with radiation modes (Fig. \ref
{Fig.12}).

In Figure \ref{Fig.13} we present the case $4B^2>1$. Here the amplitude of
radiation mode is greater than the case presented in Fig. \ref{Fig.12}.
Notwithstanding, the original shape of the kink is recovered.

The continuous spectrum (which concerns with radiation) was considered in
our calculations and numerical experiments. In the numerical experiments
there are emission of radiation that is absorbed by the resistance
(dissipative terms).

When the soliton is trapped the translational mode has a relevant role in
contrast with the case for which the free soliton enters in interaction with
an impurity. For the last case the shape modes indeed play a determinant
mode.

\section{Summary and Conclusions}

In this paper we have presented the importance of considering the soliton as
an extended particle as it interacts with real inhomogeneous media.

We have shown that a finite-width impurity can activate a large number of
soliton internal modes. In fact, for an impurity of the $N(x)\phi $ type
with a stable point the number of mode increases (as the width of the
impurity increases) in comparison with the free soliton case. Surprisingly,
this contrasts with the inhomogeneous external force $F(x)$ with a stable
point case, where the opposite occurs. An interesting finite-size effect
occurs when both inhomogeneities, $F(x)$ and $N(x)\phi $, exhibiting only
one equilibrium point when considered individually, generate a double-well
effective potencial for the soliton as they interact between each other and
with the soliton.

In addition, we have predicted surprising effects when more than one
impurity is considered. For instance, the stability condition depends on a
variety of parameters that we can control (i.e., the width and the height of
the impurity as well as the width of the soliton) and which we present as
very relevant to the whole dynamics.

We have shown the importance of the internal modes of the soliton as they
can generate shape chaos for the soliton as well as cases in which the first
shape mode leads the dynamics.

Our approach has been also applied to the stochastic resonance\cite{g:51} of
solitons that can oscillate in a double-well potential\cite{g:5}. Our
approach provides the possibility to encounter resonances linked to the
different internal modes of the soliton as well as to tune the
time-dependent driving force to a selected mode.

The results presented in this paper are very relevant for the concrete
physical systems presented in Sec. II. as they concern the pinning problem
(i.e., the stability problem of the soliton in an equilibrium point created
by the inhomogeneities). As we have seen, it is not trivial to determine
whether the equilibrium position is stable or not for the soliton when we
are in the presence of external fields and inhomogeneities having finite
widths and, also the extended character of soliton is considered. Real
experiments\cite{g:6,g:7} where this kind of phenomena has been observed
show that, in certain cases, the description of the impurities using delta
functions can lead to erroneous conclusions.

Soliton oscillators in Josephson junctions have been studied intensively in
the last years due to their application as sources of radiation\cite{g:8}.\
In this context the study presented in this paper of the sustained
oscillations of a soliton in an effective potential well has great
importance. Gr\o nbech-Jensen and Blackburn\cite{g:9} have studied a system
of coupled Josephson junctions as a super-radiant power source. Their point
is that the oscillator ceases to be a point-like oscillator and exhibits an
spatial degree of freedom. This leads to a radiation power greater than the
expected from the point-like oscillators theory. In this paper we have shown
that under the effects of the inhomogeneities the soliton exhibit an
internal spatial degree of freedom. Furthermore, there is an increased
number of activated internal modes. Particularly, we have shown that for
impurities of the type $N(x)\phi $ with a stable equilibrium position, the
hyperradiant effect could be achieved varying the impurity width. In this
case the width of soliton will be greater than the width of the unperturbed
soliton. This is also valid for impurities of the type $N(x)\sin \phi $
which can describe a microshort in a sine-Gordon-like systems. The chaotic
oscillations and the soliton explosion phenomanon (due to soliton
instability against the interaction with radiation modes) can be also
important for the design of soliton based devices.

\acknowledgments 

This work has been partially supported by Consejo Nacional de
Investigaciones Cient\'ificas y Tecnol\'ogicas (CONICIT) under Project
S1-2708.

\appendix

\section{Numerical Method}

We have integrated our $\phi ^4$-like equations using a standard implicit
finite difference method with open boundary conditions $\phi _x(0,t)=\phi
_x(l,t)=0$ and a system length $l=80$.

We use a kink-soliton as initial condition:

\begin{equation}
\phi (x,0)=A\tanh \left[ \frac{B(x-x_0)}{\sqrt{1-v_0^2}}\right]  \label{aa:1}
\end{equation}
and 
\begin{equation}
\phi _t(x,0)=\frac{-ABv_0}{\sqrt{1-v_0^2}}\cosh {}^{-2}\left[ \frac{B(x-x_0)%
}{\sqrt{1-v_0^2}}\right] ,  \label{aa:2}
\end{equation}
here $A$ and $B$ are constants whereas $x_0=1.0$ and $v_0=0$ are the initial
position and velocity of the center of mass of the soliton.

We define the position of the center of mass of the kink-soliton as,

\begin{equation}  \label{aa:3}
x_{c.m.}=\frac{\int_{-l/2}^{l/2}x\phi _x^2dx}{\int_{-l/2}^{l/2}\phi _x^2dx}.
\end{equation}

\section{The Karhunen-Lo\`eve Decomposition}

The Karhunen-Lo\`eve decomposition\cite{g:1,g:2,g:3,g:4} allows to describe
the dynamics in terms of an adequate basis of orthonormal functions or
modes. The field $\omega (x,t)$ to be decomposed represents the fluctuations
of $\phi (x,t)$ with respect to the time-averaged spatial pattern $\phi
_{ave}(x)$. We find a basis of orthonormal functions $\Psi _n(x)$ by solving
an integral equation whose kernel is the two points correlation function $%
K(x,x^{\prime })=\langle \omega (x,t)\omega (x^{\prime },t)\rangle $ (here $%
\langle ...\rangle $ means time average). The functions $\Psi _n(x)$ are the
eigenfunctions of the integral equation,

\begin{equation}  \label{bb:1}
\int_0^LK(x,x^{\prime })\Psi _n(x^{\prime })dx^{\prime }=\lambda _n\Psi
_n(x).
\end{equation}

The eigenvalues $\lambda _n$ can be regarded as the weight of the mode $n$.

\begin{figure}[tbp]
\caption{Scenario of inhomogeneous external forces. Bifurcation diagram for
the position of the center of mass of the kink soliton ($\nu = 0.26 ... 0.36 
$, $\gamma = 0.450$).}
\label{Fig.1}
\end{figure}

\begin{figure}[tbp]
\caption{Time series $x_{c.m.}$ vs $t$ for the position of the center of
mass of the kink soliton, $\gamma = 0.450$. (a) Period eight, $\nu = 0.310$.
(b) Period four, $\nu =0.312$. (c) Chaos, $\nu =0.328$. (d) Period three, $%
\nu =0.360$.}
\label{Fig.2}
\end{figure}

\begin{figure}[tbp]
\caption{Periodic, quasiperiodic and chaotic attractors, $\gamma = 0.450$.
(a) Phase space $v_{c.m.}$ vs $x_{c.m.}$, $\nu =0.322$. (b) Poincar\'e map
revealing a two-torus ($\nu =0.327$) and a strange attractor ($\nu =0.332$).
The inset presents the detail of one of the toruses.}
\label{Fig.3}
\end{figure}

\begin{figure}[tbp]
\caption{Time series $x_{c.m.}$ vs $t$ for the position of the center of
mass of the kink soliton, $\nu =0.300$. (a) Chaotic jumps of the soliton
between the two wells, $\gamma =0.505$. (b) Chaotic soliton motion mainly in
one well, $\gamma =0.550$.}
\label{Fig.4}
\end{figure}

\begin{figure}[tbp]
\caption{Chaotic solitons, $\nu =0.300$. (a) Poincar\'e map revealing a high
dimensional attractor as the soliton jumps between the two wells, $\gamma
=0.505$. (b) Low dimensional attractor for kink motion mainly in one well, $%
\gamma =0.550$. (c)-(d) Spatiotemporal evolution for previous attractors.}
\label{Fig.5}
\end{figure}

\begin{figure}[tbp]
\caption{Scenario of finite-width impurity. (a) Phase space $v_{c.m.}$ vs $%
x_{c.m.}$ for a period-three solution, $\nu =0.160$ and $\gamma =0.100$. (b)
One-torus quasiperiodic solution, $\nu =0.195$ and $\gamma =0.100$. (c)
Phase space for a three-torus quasiperiodic solution, $\nu =0.120$ and $%
\gamma =0.010$. (d) Poincar\'e map for the three-torus attractor.}
\label{Fig.6}
\end{figure}

\begin{figure}[tbp]
\caption{Scenario of finite-width impurity and inhomogeneous forces, $%
\left\langle \phi (x)\right\rangle$ vs $x$ for $\nu =0.20$ (period one), $%
\nu =0.28 ... 0.45$ (quasiperiodicity), $\nu =0.55$ (chaos) and $\nu =0.60$
(period one). $\gamma =0.300$.}
\label{Fig.7}
\end{figure}

\begin{figure}[tbp]
\caption{Scenario of finite-width impurity and inhomogeneous forces:
quasiperiodic bifurcations, $\gamma =0.300$. (a)-(b) Phase space and
Poincar\'e map for $\nu =0.280$. (c)-(d) Phase space and Poincar\'e map for $%
\nu =0.350$.}
\label{Fig.8}
\end{figure}

\begin{figure}[tbp]
\caption{Shape chaos, $\gamma =0.300$ and $\nu =0.550$. (a) High dimensional
Poincar\'e map. (b) Spatiotemporal evolution.}
\label{Fig.9}
\end{figure}

\begin{figure}[tbp]
\caption{(a) Karhunen-Lo\`eve spectra for the sequence of bifurcations
presented in Fig. \ref{Fig.7}. (b) Plot of the first mode of the
Karhunen-Lo\`eve spectrum for $\nu =0.20$ and $\nu =0.60$. The inset shows
the traslational mode and the first shape mode.}
\label{Fig.10}
\end{figure}

\begin{figure}[tbp]
\caption{(a)-(b) Spatiotemporal evolution of the period one solutions that
initiates and ends the sequence of bifurcations presented in Fig. \ref{Fig.7}%
.}
\label{Fig.11}
\end{figure}

\begin{figure}[tbp]
\caption{(a)-(j) Interaction of the soliton with radiation modes: soliton
explosion. After the explosion an anti-kink is formed. Note that the initial
kink profile is stretched because the equilibrium position is unstable
whereas the opposite occurs for the final stage. ($L=80$, $\gamma =0.1$, $%
B=0.2$ and $C=-0.001$).}
\label{Fig.12}
\end{figure}

\begin{figure}[tbp]
\caption{(a)-(b) Interaction of the soliton with radiation modes under
stability conditions for the continuous spectrum. The amplitude of the
radiation mode is greater than the case presented in Fig. \ref{Fig.12} as is
evident in the snapshot (a) for $t=0$ ($L=80$, $\gamma =0.1$, $B=0.8$ and $%
C=-0.011$).}
\label{Fig.13}
\end{figure}

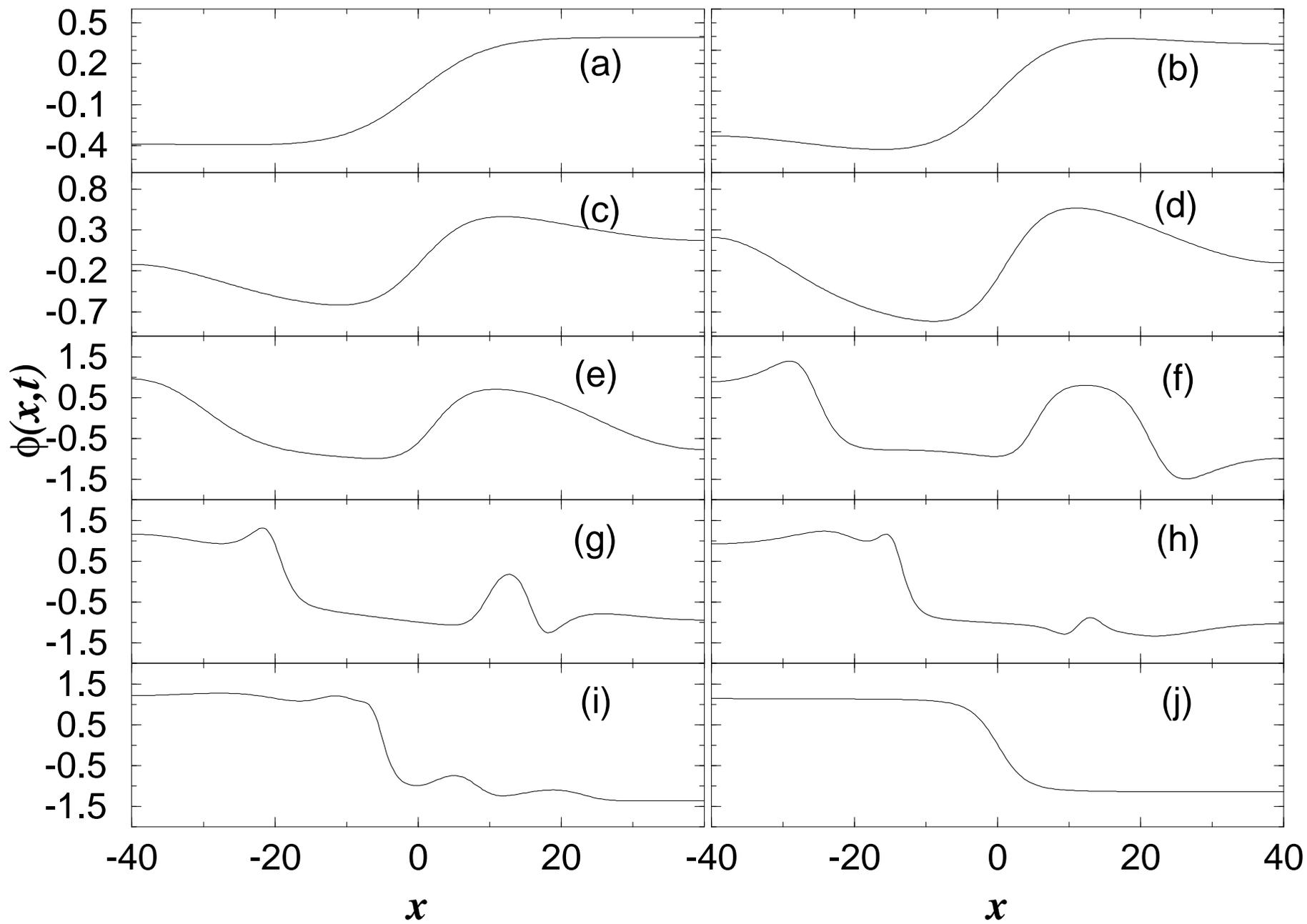
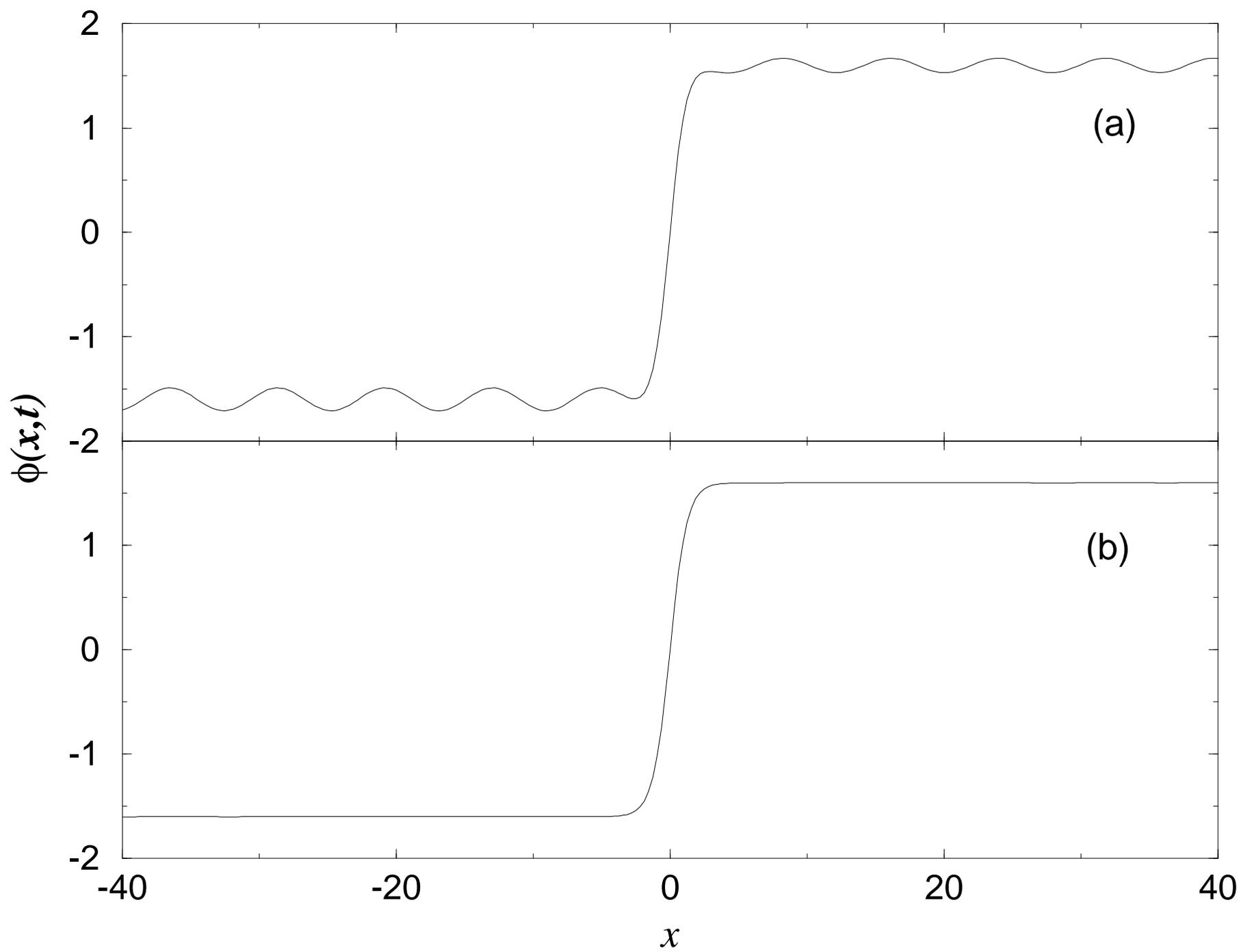
\end{document}